\documentclass[12pt,preprint]{aastex}

\newcommand{\p}{{\partial}}
\newcommand{\N}{{{N}}}
\newcommand\EQ{\begin{equation}}
\newcommand\EN{\end{equation}}
\newcommand\EQA{\begin{eqnarray}}
\newcommand\ENA{\end{eqnarray}}
\newcommand{\gapprox}{\lower.4ex\hbox{$\;\buildrel >\over{\scriptstyle\sim}\;$}}
\newcommand{\lapprox}{\lower.4ex\hbox{$\;\buildrel <\over{\scriptstyle\sim}\;$}}

\shorttitle{Ambipolar}
\shortauthors{Leprovost \& Kim}

\begin{document}

\title{SELF-CONSISTENT MEAN FIELD THEORY \\
IN WEAKLY IONIZED GAS}

\author{Nicolas Leprovost}
\affil{Groupe Instabilit\'e et Turbulence, CEA/DSM/DRECAM/SPEC,
F-91191 Gif sur Yvette Cedex, France}
\email{nicolas.leprovost@cea.fr}

\and

\author{Eun-jin Kim}
\affil{Department of physics, University of California, San Diego, La Jolla, CA 92093-0319, USA}
\email{ejk@physics.ucsd.edu}

\begin{abstract}
We present a self-consistent mean field theory of the dynamo
in 3D and turbulent diffusion in 2D in weakly ionized gas.
We find that in 3D,
the backreaction does not alter the beta effect while it suppresses
the alpha effect when the strength of a mean magnetic field
exceeds the critical value  $B_c \sim\sqrt{\nu_{in} \tau_n \langle {v}^2 \rangle /R_m}$.
Here, $\nu_{in}$, $\tau_n$, and $R_m$ are ion-neutral collision frequency,
correlation time of neutrals, and magnetic Reynolds number, respectively.
These results suggest that a mean field dynamo operates much more
efficiently in weakly ionized gas where $\nu_{in} \tau_n \gg 1$,
compared to the fully ionized gas.
Furthermore, we show that in 2D, the turbulent diffusion is
suppressed by back reaction when a mean magnetic field
reaches the same critical strength $B_c$,
with the upper bound on turbulent diffusion given by its kinematic
value. Astrophysical implications are discussed.
\end{abstract}


\section{Introduction}

One of the most outstanding problems in the astrophysical MHD is
to explain the origin of ubiquitous magnetic fields in stars,
galaxies, interstellar medium (ISM), etc. These magnetic
fields are often observed to be coherent on scales
much larger than the characteristic scale of turbulence, with
their energy being comparable with fluid kinetic energy
(i.e., in equipartition).
For instance, galactic magnetic fields are thought to 
have coherent magnetic fields with comparable strength to fluctuations.
The major stumbling block to explaining these coherent
magnetic fields by dynamo action in fully ionized gas is its tendency of
generating too strong fluctuations, which unfortunately
inhibit the growth of a coherent (mean) component by
back reaction (Lorentz force) when the strength of a mean magnetic
field is far below equipartition value -- the so--called
alpha quenching problem \citep{Cattaneo96,Gruzinov94,Gruzinov96}.

It is, however, largely unknown whether and/or how backreaction constrains
alpha effect in weakly ionized medium,
such as the galaxies, ISM, molecular clouds, etc,
with ambipolar drift (slippage between magnetic fields and the bulk
of fluid (neutrals)).
This is partly because almost all previous works on the effect of
ambipolar drift invoked strong coupling approximation
(the drift between ions and neutrals is balanced by Lorentz force
due to sufficiently frequent collisions between the two),
%, i.e., ${\bf v}_i - {\bf v}_n = ?$),
which makes ambipolar drift
mainly act as a nonlinear diffusion. Thus, the ambipolar drift
has been primarily advocated as a means of enhancing the effective
diffusion rate over Ohmic value \citep[e.g.,][]{Mestel56,Zweibel88}.
It is also attributed to the fact that the important dynamic effect
of fluctuations (turbulence and Lorentz force back reaction)
has often been neglected \citep[c.f.,][]{Boss00,Fatuzzo02}.
Interestingly, these two factors come in together as strong coupling
approximation is likely to break down on small scales (i.e., for
fluctuations) where Alfven frequency is larger than
the ion-neutral collision frequency \citep{Kim97}.
The purpose of this Letter is to
present a first self-consistent mean-field
theory of the dynamo in weakly ionized gas, by
incorporating these important synergistic effects of
turbulence and back-reaction, without invoking strong coupling 
approximation.
We shall demonstrate that ambipolar drift reduces alpha quenching, even
overcoming it in certain (but extreme) cases.

Before proceeding to a mean field theory of the dynamo,
some insight into the effect of ambipolar drift can be gained by
considering the problem of diffusion of a mean magnetic
field in the two dimensions (2D).
In the case of fully ionized gas, it is well known
that  the turbulent diffusion in 2D is severely reduced by back reaction
\citep{Cattaneo91,Gruzinov94}.
In weakly ionized gas, the turbulent diffusion is
still reduced below its kinematic value while
ambipolar drift can increase the critical strength of a mean
magnetic field (above which the diffusion is
reduced) by a factor of $\sqrt{\nu_{in}\tau_n}$ \citep{Kim97}.
Here, $\nu_{in}$ and $\tau_n$
are the ion-neutral collision frequency and correlation time of neutrals.
Thus, ambipolar drift offers the possibility of dissipating
a mean magnetic field at turbulent rate for sufficiently large 
$\nu_{in}\tau_n$.
In this Letter, we also provide a self-consistent mean field theory
for the diffusion of a mean magnetic field in 2D in weakly ionized gas,
which not only confirms the numerical results of \citet{Kim97} but
also generalizes the perturbation analysis therein.

The remainder of the Letter is organized as follows.
Section 2 contains the derivation of the effective dissipation rate
of a mean magnetic field with ambipolar drift in 2D.
We provide a mean field dynamo theory in weakly
ionized gas in \S3, by deriving the analytic expressions for alpha and 
beta effects.
Concluding remarks are provided in \S 4.

\section{TURBULENT DIFFUSION OF MEAN MAGNETIC FIELD IN 2D}
%\label{2D}

In weakly ionized medium, Ohm's law is valid provided that
the velocity of ions is used in the calculation of current.
So, to consistently treat this problem, one needs to solve the momentum
equation for ions and for neutrals (without Lorentz Force),
together with induction equation for magnetic fields.
Since $\nu_{in}/{\rho_n}= \nu_{ni}/{\rho_i}$ and
$\rho_i/\rho_n \ll 1$ for weakly ionized medium,
the neutral-ion collision frequency can be neglected.
Here, $\rho_n$ and $\rho_i$ are the density of neutrals and  ions, 
respectively,
and $\nu_{in}$ and $\nu_{ni}$ are ion-neutral and neutral-ion  collision 
frequency,
respectively.
Thus, the momentum equation for the neutrals is entirely
decoupled from that of the ions as well as from the induction equation.
We thus assume that neutral velocity is turbulent with a prescribed
statistics, and then solve the momentum equation for ions,
which evolves self-consistently by frictional coupling to neutrals
and by Lorentz force.
Note that we are not invoking the strong coupling approximation.

In 2D, we work with ion vorticity $\omega$
($\omega \hat{z} = \nabla \times {\bf v}$) and magnetic
potential $A$ (${\bf B}=\nabla \times (A \hat{z})$), which
are governed by the following set of two equations (in dimensionless form):
\begin{eqnarray}
(\p_t + {\bf v} \cdot \nabla)\omega &=& -({\bf B} \cdot {\nabla}) 
{\nabla}^2 A + \chi {\nabla}^2 \omega + \gamma ({\N}-\omega) \,,
\label{syst2D} \\
(\p_t + {\bf v} \cdot \nabla)A &=& \eta {\nabla}^2 A\,.
\nonumber
\end{eqnarray}
Here, ${\bf v}$ is the ion velocity,
${\N}$ is the vorticity of neutrals, $\gamma=\nu_{in} \tau_n$ is the 
frictional
coupling between ions and neutrals, $\eta$ is Ohmic diffusivity,
and $\chi$ is ion viscosity. We shall assume unity magnetic Prandtl
number (i.e., $\eta = \chi$).
By decomposing fields into  large and small scale
parts and assuming that there is no large scale
displacement of the medium, we let
${\bf v}={\bf v_0} +  {\bf v} =  {\bf v}, \omega=\omega_0 +  \omega =  
\omega,
N=N_0 +  N =  N, {\bf B}={\bf B_0} +  {\bf b}$, and $A=A_0 +  a$.
Here, subscript `$0$' denotes a  mean component, averaged over the
statistics of $N$.
Using this decomposition, we can separate the system eq.\ (\ref{syst2D}) in
large and small scale components to obtain
\EQA
\label{Systeme}
[\partial_t - \chi {\nabla}^2 +\gamma] \omega &=& -({\bf B_0} \cdot {\bf
\nabla}) {\nabla}^2 a + \gamma  N \,, \\ \nonumber
[\partial_t - \eta {\nabla}^2] A_0 &=& - \langle {\bf v} \cdot {\bf 
\nabla}  a \rangle
= - {\bf \nabla} \cdot {\bf G}\,, \\ \nonumber
[\partial_t - \eta {\nabla}^2]  a &=& -  {\bf v} \cdot {\bf \nabla} A_0\,.
\ENA
Here, angular brackets denote the average over the statistics of $N$; in 
the first
equation, the term ${\nabla}^2 A_0$ was neglected due to the large-scale 
variation of
$A_0$;  in the last equation, the term $ {\bf v}
\cdot {\bf \nabla}  a  - \langle {\bf v} \cdot {\bf \nabla} 
a \rangle$ was dropped (quasi linear approximation).
${\bf G} = \langle {\bf v} a \rangle$ is the flux of magnetic potential,
which determines the
evolution (effective diffusion) of $A_0$. To obtain ${\bf G}$ in terms of
mean quantities, we rewrite it as:
\EQ
{\bf G}=\langle {\bf v} \int dt \partial_t  a \rangle + \langle \int dt 
\partial_t 
{\bf v} a \rangle = {\bf G}_1 + {\bf G}_2\,.
\EN
Here, ${\bf G}_1$ is a kinematic part
while ${\bf G}_2$ comes from the back reaction of the flow onto
the magnetic potential. It is easy to check ${\bf G}_1 = - \frac{\tau}{2}
\langle  {v}^2 \rangle {\bf \nabla} A_0$,  by using $\tau$ approximation
\citep{Gruzinov94,Gruzinov96}, namely by replacing
the time derivative by $\frac{1}{\tau}$.
The expression for ${\bf G}_1$ is just the standard beta effect in
2D. It is interesting to express ${\bf G}_1$ in terms of
$N$ since the statistics of the latter can be prescribed.  For
simplicity, we assume the statistics of $N$ to be
stationary with a delta-function power spectrum around
$k=k_0$ as follows:
\EQ
\label{statistics}
\langle N({\bf k_1},t) N({\bf k_2},t)\rangle=
\frac{\langle {N}^2 \rangle }{2\pi k_0}\delta({\bf
k_1}+{\bf k_2})\delta(k_1-k_0)\,.
\EN
By taking spatial Fourier transform of the first equation of eq.\ (\ref{Systeme})
without the Lorentz force term,
and by using eq.\ (\ref{statistics}), we obtain
\EQ
\langle {v}^2 \rangle = [\frac{\tau \gamma}{1+\tau \gamma}]^2 \frac{\langle 
N^2 \rangle}{k_0^2}\,.
\EN
Therefore, the effective diffusion coefficient in the kinematic limit is 
given by
$\beta_0= - \frac{{\bf G_1}}{{\bf \nabla} A_0}= \frac{\tau}{2} [\frac{\tau
\gamma}{1+\tau \gamma}]^2 \frac{\langle N^2 \rangle}{k_0^2}$.
Note that $\beta_0$ takes its maximum value
when neutrals and ions are strongly coupled with $\gamma \tau \gg  1$.
This is a natural consequence of the assumption that ions obtain their
kinetic energy through frictional coupling to neutrals.
Thus, crudely put, $\beta_0$ is reduced by a factor
$[\frac{1+\tau \gamma}{\tau \gamma} ]^2$.
If there were an independent
energy source for ions, this would no longer be true.

To compute ${\bf G}_2$, we incorporate Lorentz force in the first equation
of eq.\ (\ref{Systeme}) and take the Fourier transform to obtain:
\EQ
\tilde{\omega}({\bf k}) = \frac{\tau}{1+\tau \gamma} [-\varepsilon_{ij3} 
\int d{\bf k'} ({\bf k} -{\bf k'})_j \tilde{A_0}({\bf k} -{\bf k'}) k'_i 
k'^2 \tilde{a}({\bf k'}) + \gamma \tilde{N}({\bf k})] \,,
\EN     
from which ${\bf G}_2$ follows as:
\EQA
\label{equav2}
{\bf G}_2 &=& - i \int d{\bf k} d{\bf k'} e^{[i({\bf k}+{\bf 
k})\cdot{\bf x}]} \frac{k_l}{k^2} \varepsilon_{3lm} \langle 
\tilde{\omega}({\bf k}) \tilde{a} ({\bf k'}) \rangle  \\ \nonumber
&=& - \frac{\tau}{2(1+\tau \gamma)} \langle a {\nabla}^2 a \rangle {\bf 
\nabla} A_0 - \frac{i \tau \gamma}{1+\tau \gamma} \int d{\bf k} d{\bf 
k'} e^{[i({\bf k}+{\bf k})\cdot{\bf x}]} \frac{k_l}{k^2} 
\varepsilon_{3lm} \langle \tilde{N}({\bf k}) \tilde{a}({\bf k'}) \rangle \,.
\ENA
To calculate the first part of the RHS, we assumed that the magnetic 
potential fluctuations were isotropic and homogeneous. 
Since the neutrals are unlikely to be correlated with the magnetic field,
the second part of the preceding equation can be neglected, leading to
\EQ
\label{eqbeta}
\beta=\beta_0+\frac{\tau}{2(1+\tau\gamma)}\langle a {\nabla}^2 a \rangle
=\beta_0-\frac{\tau}{2(1+\tau\gamma)}\langle b^2 \rangle\,.
\EN
Compared to $\beta = \tau \langle v^2 - b^2 \rangle /2$ in the fully
ionized gas, the contribution from the backreaction in eq.\ (\ref{eqbeta})
involves a multiplicative factor $1/(1+ \tau \gamma)$. It is  because
the response of ions and magnetic field are different due to frictional
coupling of ions to neutrals (see eq.\ (\ref{Systeme})).
To express $\langle a {\nabla}^2 a \rangle$ in terms of large scale 
quantities,
we use Zeldovich theorem \citep{Zeldovich57},
which can be derived from
the conservation of $\langle A^2 \rangle$ in 2D-ideal MHD (by
multiplying the third equation of eq.\ (\ref{Systeme}) by $ A$ and
taking average over large scales) as
\EQ
\label{zeldovich}
\eta \langle a {\nabla}^2 a \rangle=\langle a  {\bf v}\rangle
 \cdot {\bf \nabla}A_0=-\beta({\bf
\nabla}A_0)^2\,.
\EN
Thus, from eqs.\ (\ref{eqbeta}) and (\ref{zeldovich}), we obtain
$\partial_t A_0 =  (\eta+\beta) {\nabla}^2 A_0$ with
\EQA
\beta&=&\frac{\beta_0}{1+\frac{\tau}{\eta  (1+\gamma \tau)}({\bf
\nabla}A_0)^2}\,.
\ENA
Note that $(\eta+\beta)$ is the total effective diffusivity of $A_0$.
In the case of weak coupling limit ($\tau \gamma \ll 1$), the previous 
equation
reduces to the beta
suppression in fully ionized gas for a given $\beta_0$. Note, however,
that $\beta_0$ itself is proportional to $(\tau \gamma)^2/(1+\tau \gamma)^2
\sim (\tau \gamma)^2$.
In the opposite strong coupling limit ($\tau \gamma \gg 1$), 
one recovers an expression similar to that of \citep{Kim97} as
$\beta \sim \beta_0/(1+B_0^2/\eta \gamma)$. Thus,
back-reaction becomes insignificant when large scale magnetic
field is weak enough so as to satisfy  the following condition:
\EQ
\label{condition}
({\bf \nabla}A_0)^2 = B_0^2 \ll \eta \gamma
=  {\gamma\over R_m} \langle v^2 \rangle\,,
\EN
where $R_m = \tau \langle {v}^2 \rangle / \eta$ is the magnetic Reynolds 
number.
Thus, the critical strength of mean magnetic
field for the suppression of $\beta$ effect is
$\gamma \langle v^2 \rangle/ R_m$,  which is larger by a factor
of $\gamma$ than that in the fully ionized gas.   
Note that the turbulent diffusivity can reach its kinematic value $\beta_0
= \tau \langle {v}^2\rangle /2$ as $\gamma \tau_n \to \infty$
but can never be greater. That is, $\beta_0$ is the {\it upper limit}
on $\beta$.
These results agree with \citet{Kim97}.

\section{MEAN FIELD DYNAMO THEORY IN 3D}
\label{3D}
We now provide a mean field dynamo theory in weakly ionized
gas in 3D,  by self-consistently computing the alpha and beta effects.
As previously, we use quasi-linear theory
to obtain the following set of non-dimensionalized equations for 
fluctuations and
mean field (denoted by a subscript `0'):
\EQA
\label{Syst3D}
[\partial_t+\gamma-\chi{\nabla}^2]{\bf v}&=&{\bf B}_0\cdot{\bf \nabla}
{\bf b} + {\bf b}\cdot{\bf \nabla} {\bf B}_0 -{\bf \nabla}p+\gamma {\bf
{\bf N}}\,,\\ \nonumber
[\partial_t-\eta {\nabla}^2]{\bf b}&=&{\bf \nabla} \times ({\bf v} \times
{\bf B}_0)\,,\\ \nonumber
[\partial_t-\eta {\nabla}^2] {\bf B}_0&=& {\bf \nabla} \times \langle {\bf v}
\times {\bf b}\rangle = {\bf \nabla} \times {\bf E}\,.
\ENA
Here, ${\bf N}$ is neutral velocity (not vorticity);
${\bf E}=\langle {\bf v} \times {\bf b}\rangle = \alpha {\bf B}_0 - 
\beta \nabla \times
{\bf B}_0$
is the electromotive force, which contains $\alpha$ and $\beta$ effects.
To compute ${\bf E}$, we again consider 
two parts -- the kinematic part ${\bf E}_0$ and the part coming from the 
back
reaction of the magnetic field onto the fluids ${\bf E}_1$:
\EQA
\label{ElecForce}
{\bf E}&=&\langle  {\bf v} \times \int dt \partial_t  {\bf v} \rangle  + 
\langle \int dt
\partial_t  {\bf v} \times {\bf b}\rangle  \\ \nonumber 
&=& \alpha_0 {\bf B}_0 - \beta_0 {\bf \nabla} \times {\bf B}_0 + \langle 
\int
dt \partial_t  {\bf v} \times {\bf b}\rangle  = {\bf E}_0 + {\bf E}_1\,,
\ENA
where $\alpha_0 = - \frac{\tau}{3}\langle {\bf v}\cdot {\bf \nabla} 
\times {\bf
v}\rangle $ and $\beta_0 =  \frac{\tau}{3}\langle v^2\rangle$ are the 
kinematic values
\citep[c.F.,][]{Krause80}. These two coefficients can again be 
expressed in
terms of ${\bf N}$ as:
\EQA
\label{alphakin}
\alpha_0 &=& -\frac{\tau}{3}[\frac{\gamma \tau}{1+\tau\gamma}]^2\langle 
{\bf
N}\cdot{\bf \nabla}\times{\bf N}\rangle\,, \\ \label{betakin}
\beta_0 &=& \frac{\tau}{3}[\frac{\gamma \tau}{1+\tau \gamma}]^2 \langle 
N^2\rangle \,.
\ENA

%STATISTICS AND DERIVING SIMULTANEOUSLY $\alpha$ AND $\beta$ SUPPRESSION}
%\medskip

The computation of ${\bf E}_1$ can most easily be done in Fourier space
because of the pressure term.
Thus, we write the equation for the velocity in Fourier space and then 
plug it
into the electromotive force expression to obtain:
\EQA
\label{EFourier}
E_{1\alpha}&=&\frac{i \tau}{1+\tau \gamma} \; \epsilon_{\alpha \beta
\gamma} \int d{\bf p} \int d{\bf q} \; \Gamma_{\beta \lambda \mu}({\bf
k}-{\bf p}) B_{0\lambda}({\bf q})\langle b_{\mu}({\bf k}-{\bf p}-{\bf
q})b_{\gamma}({\bf p})\rangle \,,\\ \nonumber
\Gamma_{\alpha \beta \gamma}({\bf k})&=&\delta_{\alpha \beta} k_{\gamma}
+ \delta_{\alpha \gamma} k_{\beta}-2k_{\alpha} k_{\beta} k_{\gamma}/{k}^2\,.
\ENA
To compute $E_{1\alpha}$, we assume that the statistics of small scale 
magnetic fields
is homogeneous and isotropic but not necessarily invariant under plane 
reflection,
with the following correlation function:
\EQ
\label{StatB}
\langle b_{\alpha}({\bf k})b_{\beta}({\bf k'})\rangle  =\delta({\bf k}+{\bf
k'})[\frac{M(k)}{4\pi {k}^2}(\delta_{\alpha \beta}-k_\alpha k_\beta/{k}^2) +
i \frac{F(k)}{8\pi {k}^4} \epsilon_{\alpha \beta \gamma}
k_\gamma]=\delta({\bf k}+{\bf k'})\Phi_{\alpha \beta}({\bf k})\,,
\EN
where $M(k)$ is the magnetic energy spectrum tensor and $F(k)$ is the
magnetic helicity spectrum tensor. By using eq.\ (\ref{StatB}) in (\ref{EFourier})
and by keeping terms up to $k$ (stretching and diffusion term), we obtain:
\EQ
E_{1\alpha}=\frac{i\tau}{1+\tau \gamma} \; \epsilon_{\alpha \beta
\gamma} B_{0\lambda}({\bf k}) \int d{\bf p} \; \Phi_{\mu \gamma}({\bf
p})[(2\frac{p_\beta p_\lambda p_\mu}{{p}^2} - \delta_{\beta \lambda} p_\mu
- \delta_{\beta \mu} p_\lambda) + (\delta_{\beta \lambda}-\frac{2
p_\beta p_\lambda}{{p}^2})k_\mu]\,.
\EN
Since all integrals with odd numbers of $p_i$
vanish, $E_{1\alpha}$ reduces to:
\EQA
E_{1\alpha}=\frac{i\tau}{1+\tau \gamma} \; \epsilon_{\alpha \beta
\gamma} B_{0\lambda}({\bf k})[k_\mu \int d{\bf p} \; \frac{M(p)}{4\pi
p2}(\delta_{\mu \gamma}-p_\mu p_\gamma/p^2) (\delta_{\beta
\lambda}-\frac{2 p_\beta p_\lambda}{p^2})\\ \nonumber
+ i \int d{\bf p} \; \frac{F(p)}{8\pi p4} \epsilon_{\mu \gamma \delta}
p_\delta (2\frac{p_\beta p_\lambda p_\mu}{p^2} - \delta_{\beta \lambda}
p_\mu - \delta_{\beta \mu} p_\lambda)] \,.
\ENA
The first part (proportional to $k_\mu$), contributing to $\beta$, 
vanishes when
integrated over angles, while the second part gives the correction term to
$\alpha$ due
to back reaction. Thus, there is no change in $\beta$,  namely, {\it the 
turbulent diffusion is not
affected by the back reaction of small scale magnetic fields in 3D
with ambipolar drift!} This result sharply contrasts to the claim
made in the literature, based on strong coupling approximation,
that ambipolar drift enhances the diffusion of a mean magnetic field
in 3D \citep[e.g.,][]{Subramanian98}. 
Note that the result for fully ionized gas \citep{Gruzinov96}
is recovered simply by taking the limit $\tau \gamma \to 0$, but by
keeping $\beta_0$ constant.
On the other hand,
the surviving part of $E_1$, contributing to $\alpha$, reads:
\EQ
\label{Efinal}
E_{1\alpha} = \frac{\tau}{1+\tau \gamma} \frac{\langle {\bf b}\cdot{\bf 
\nabla}
\times {\bf b}\rangle }{3} B_{0\alpha}\,.
\EN
Therefore, $\alpha$ effect, including the back reaction of the magnetic 
field, follows
from eqs.\ (\ref{ElecForce}), (\ref{alphakin}), and (\ref{Efinal}) as
\EQ
\label{alpha1}
\alpha=\alpha_0+\frac{\tau}{1+\tau \gamma} \frac{\langle {\bf b}\cdot{\bf
\nabla} \times {\bf b}\rangle }{3}\,.
\EN
Note that only the helical (resp. non-helical) part of the magnetic spectrum is 
involved in the alpha (resp. beta) effect since $\alpha$ (resp. $\beta$) is a pseudo-scalar
(resp. scalar). 
Compared to the fully ionized gas, the contribution from the current 
helicity to $\alpha$ contains the additional multiplicative
factor of $1/(1+ \tau \gamma)$. This is again because the response
time of ions is different from that of magnetic
fields due to frictional coupling to neutrals (see eq.\ (\ref{Syst3D})).
Thus, it is very likely that the cancellation between fluid and current 
helicity
for Alfven waves (as happens in fully ionized case with $\gamma=0$)
may be avoided for $\gamma \tau >1$, thereby reducing the suppression
of $\alpha$ effect. This shall be shown below. To close the expression
for $\alpha$, we need to express the current helicity in terms of mean  
magnetic
fields. To do so, we use the topological invariant of mean magnetic helicity
$\langle {\bf a} \cdot {\bf b}\rangle $
in 3D, from which an analog of Zeldovich theorem can be derived as
\EQ
\label{Zeldovich3D}
\eta \langle {\bf b} \cdot {\bf \nabla} \times {\bf b}\rangle =-\langle 
{\bf v} \times {\bf
b}\rangle  \cdot {\bf B_0} = -\alpha B_0^2 + \beta_0 {\bf B_0} \cdot {\bf
\nabla} \times {\bf B_0}\,.
\EN
Finally, combining eqs.\ (\ref{alpha1}) and (\ref{Zeldovich3D}), we obtain the
non-linear $\alpha$-effect expression for 3D-MHD with ambipolar drift:
\EQ
\alpha=\frac{ \alpha_0 +
\frac{\tau \beta_0}{3 \eta(1+\tau\gamma)}{\bf B_0} \cdot {\bf \nabla} 
\times
{\bf B_0}}{1+\frac{\tau}{3(1+\tau\gamma)}\frac{B_0^2}{\eta}}\,.
\EN
The previous equation recovers $\alpha$ in the case of fully ionized gas
as $\gamma \tau \to 0$. In the strong coupling limit ($\gamma \tau \gg 1$),
$\alpha$ effect is suppressed when eq.\ (\ref{condition}) is satisfied. Therefore,
the critical strength of mean magnetic field for the suppression of
$\alpha$ effect is $ \gamma \langle {v}^2 \rangle /R_m$, 
larger by a factor of $\gamma$, compared to
the case of fully ionized gas.
This has significant implications for a mean field dynamo
in the galaxy, ISM, etc where the bulk of fluid consists of neutrals
with $\gamma \gg 1$
(see \S 4 for more discussion).

\section{CONCLUSION}
We have presented self-consistent mean field theory of
the turbulent diffusion (in 2D) and
the dynamo (in 3D) in weakly ionized gas, by incorporating
turbulence and back reaction of fluctuating magnetic fields.
The key results are  that in 3D,
the backreaction does not alter the beta effect while it suppresses
the alpha effect when the strength of a mean magnetic field
exceeds the critical value  $B_c^2 \sim\gamma \langle {v}^2\rangle /R_m$.
This critical value  is larger than that $B_c^2 \sim  \langle {v}^2\rangle /R_m$
in the case of the fully ionized gas for $\gamma =\nu_{in} \tau_n >1$.
Alternatively put, the suppression factor for the alpha effect is reduced
by a factor of $\gamma$, compared to the fully ionized gas.
The upper bound on $\alpha$ is given by its kinematic value $\alpha_0
= -\tau \langle {\bf v} \cdot (\nabla \times {\bf v})/3$.
In 2D, the turbulent diffusion ($\beta$ effect) was shown to be 
suppressed by
back reaction when a mean magnetic field
reaches the same critical value $B_c^2 \sim \gamma \langle {v}^2\rangle 
/R_m$,
with the upper bound on turbulent diffusion given by the kinematic
value $\beta_0 = \tau\langle {v}^2 \rangle /2$.
These results are consistent with those in \citet{Kim97}.

Therefore, in weakly ionized gas, the degree of alpha quenching
(in 3D) and the suppression of turbulent diffusion (in 2D) crucially
depends on $\gamma = \nu_{in} \tau _n$ in addition to $R_m$,
i.e., the property of medium such as ionization, turbulence, etc.
As $\nu_{in} \sim  10^{-2} n_n$ cm$^3$/yr \citep[e.g., see][]{Kim97},
$\nu_{in} \tau_n \sim 10^{5} n_n$ for $\tau_n \sim 10^{7}$yr.
Here, $n_n$ is the number density of neutrals in unit of cm$^{-3}$.
Therefore, in the limit of a very low ionization,
the alpha quenching (and beta quenching in 2D) can be significantly 
reduced.
For instance, in the case of young galaxy with
$n_n \sim 1$cm$^{-3}$, $L \sim 100$pc, $v \sim 10$km/s, and $T \sim 10^{4}$K,
$\eta = 10^{7} (T/10^{4})^{-3/2}$cm$^2$/s $\sim 10^{7}$cm$^2$/s,
and $R_m \sim v L/\eta \sim 10^{19}$. Thus,
$\nu_{in} \tau_n/R_m \sim 10^{5}/R_m \sim 10^{-14}$, with the
critical strength of mean field $B_c \sim 10^{-7}\times
\sqrt{\langle v^2 \rangle}$, which is too weak.
However, for dark molecular clouds with $n_n \sim 10^{7}$cm$^{-3}$, $L 
\sim 1$pc,
$v \sim 1$km/s, and $T \sim 10$K,
$\eta =  3\times 10^{11}$cm$^2$/s and
$R_m \sim v L/\eta \sim 10^{12}$. Thus,
$\nu_{in} \tau_n/R_m \sim 1$, giving 
$B_c \sim \sqrt{\langle v^2 \rangle}$!
Therefore, in this extremely low ionized gas, a mean field dynamo
may work efficiently without alpha quenching.

These results essentially come from the fact that the turbulence
in weakly ionized gas does not become Alfvenic as the motion of ions 
undergoes
frictional damping due to the coupling to neutrals (or, due to ambipolar
drift). This is quite similar to what happens in a very viscous
fluid with $R_e \ll R_m$ ($R_e$ is the Reynolds number), in which
case the suppression factor for alpha quenching is also reduced
because of viscous damping of ion velocity \citep{Kim99}.
Since $R_e \ll R_m$ in the galaxies with a low ionization,
the combined effect of ambipolar drift and viscous damping of fluid
may render the mean dynamo sufficiently efficient, without
severe alpha quenching.
This interesting problem will be investigated in a future paper.

\vskip1cm

We thank P.H. Diamond and B. Dubrulle for useful comments.
N.L. is supported by programme national de chimie athmosph\'erique, and 
E.K. by the U.S. Department of Energy under Grant No.
FG03-88ER 53275.

%\bibliography{Bib_ambipolar}
%\bibliographystyle{alpha}

\end{document}